\documentstyle[prd,aps,epsf]{revtex}
\begin{document}
\twocolumn
\title{Late-Time Dynamics of Scalar Fields on Rotating Black Hole 
       Backgrounds}
\author{William Krivan}
\address{Department of Physics, University of Utah,
         Salt Lake City, UT 84112 \\
    and Institut f\"ur Astronomie und Astrophysik, Universit\"at 
        T\"ubingen, D-72076 T\"ubingen, Germany}
\maketitle

\begin{abstract}
Motivated by results of recent analytic studies, we present a numerical 
investigation of the late-time dynamics of scalar test fields 
on Kerr backgrounds. We pay particular attention to the issue of mixing of 
different multipoles and their fall-off behavior at late times.
Confining ourselves to the special case of axisymmetric modes with equatorial
symmetry, we show that, in agreement with the results of previous work, the 
late-time behavior is dominated by the lowest allowed $l$-multipole. However 
the numerical results imply that, in general, the late-time fall-off of the 
dominating multipole is different from that in the Schwarzschild case, and 
seems to be incompatible with a result of a recently published analytic 
study.
\end{abstract}

\pacs{04.30.Nk, 04.25.Dm, 04.70.Bw}

The purpose of the studies presented in this paper is to point out the need 
for further work towards a definitive answer to the following question: How 
do perturbations of a rotating black hole get radiated away?

Using the Teukolsky formalism \cite{teuk72}, we can search for an answer 
to that 
question by studying the evolution of a complex spin-weighted 
wavefunction that is defined in terms of curvature perturbations of a 
fixed black hole background geometry. 
The late-time dynamics of gravitational perturbations is governed by
features very similar to those occurring for scalar test-fields 
\cite{price72,Nils97}. 
In the special case of vanishing spin-weight, the Teukolsky 
equation reduces to the linear wave equation for a scalar test 
field $\Phi$ on a Kerr background, which has a somewhat simpler structure 
than the Teukolsky equation.
For the sake of simplicity, the present discussion is confined to the 
evolution of scalar test fields.

\begin{figure}
\epsfxsize=0.48\textwidth
\epsfbox{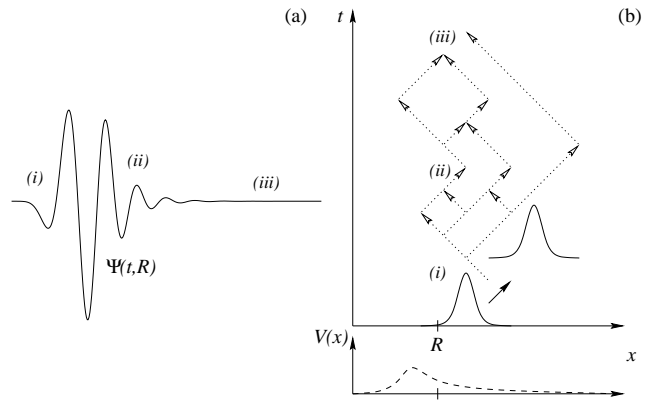}
\caption{\label{figure-1}
Wave propagation on a curved background: (a) Three stages of the dynamics: 
initial burst phase ($i$), quasinormal ringing ($ii$) and tail phase 
($iii$). (b) The quasinormal ringing and the 
tail phase are understood in terms of scattering and interference -- 
indicated by the dotted lines  -- of the wave off an effective 
potential $V(x)$, depicted by the dashed curve.
}
\end{figure}

For a spherically symmetric, asymptotically flat background geometry, the 
angular variables can be separated off, and with an appropriate radial 
coordinate $x$ the wave equation for a scalar test field $\Phi$ reduces to a $
(1+1)$ dimensional wave equation with an effective potential $V(x)$,
\begin{equation}
\label{1dwaveeq}
\left[ \partial_{tt
} - \partial_{xx
} + V(x) \right] \Psi(t,x) =0 \; .
\end{equation}
When observed from a fixed location $x=R$, the dynamics of the wave consists 
of three stages (cf.\ Fig.\ \ref{figure-1}): During the initial burst phase 
($i$), the waveform is 
dominated by the initial data, in our case given by an outgoing pulse with 
compact support. The initial burst is followed by the 
exponentially damped quasinormal ringing ($ii$) and the so-called tail phase 
($iii$), in which the amplitude of the field $\Phi$ falls off according to
a power-law, $| \Phi | \propto t^{-\mu}$.

The discovery that waves propagating on a curved background die off in 
tails goes back to the original work by R.\ Price in the early 1970s 
\cite{price72}. In the special case of a background geometry given by a 
nonrotating black hole, and when the field is not initially static,
the late time decay at a fixed location is given by $|\Phi | \propto 
t^{-(2l+3)}$, 
where $l$ is an angular separation parameter. In the last three decades, 
power-law tails on spherically symmetric backgrounds have been studied 
extensively by both analytic and numerical methods 
\cite{leaver86,gpp94a,waimo95a,Nils97,BarackII}, not only at a fixed location 
(i.e.\ at timelike infinity), but also at the horizon of the black hole and 
at future null infinity.

This situation is in dramatic contrast to the case of a background given by 
a rotating Kerr black hole, in which until very recently no analytic 
predictions for the tail behavior were available. In fact it had been quite 
unclear whether 
tails are present at all, and, if they are, what the value of the 
power-law coefficient is. Numerical work, however, was done in an attempt
to generalize the above results to the Kerr case \cite{PaperI}. 
Due to the breakdown of spherical symmetry of the background geometry, 
a multipole decomposition in terms of spherical harmonics loses its special 
meaning and it is no longer possible to expand $\Phi$ in spherical harmonics 
in the usual way. Although the functions $Y_{l}^{m}$ still form a complete 
set of functions for the description of a generic angular dependence, one 
now has to take into account the interaction between different $l$-multipoles 
in the course of the time evolution. Modes with different values of $m$, 
however, are not coupled because the background geometry is axisymmetric.
Thus in order to take into account the contributions of all the different 
multipoles occurring in the numerical time evolution, only the azimuthal 
angular 
coordinate is separated off, and one has to solve a
$(2+1)$ dimensional evolution 
problem for the field $\Psi$, defined by
\begin{equation}
\Phi \equiv \Psi(t,r^{*},\theta) e^{im\tilde{\phi}} \; ,
\end{equation}
where $r^{*}$ is the Kerr tortoise coordinate and $\tilde{\phi}$ is the 
azimuthal Kerr coordinate, that was chosen for technical reasons described in
\cite{PaperI}. 

\begin{figure}
\epsfxsize=0.48\textwidth
\epsfbox{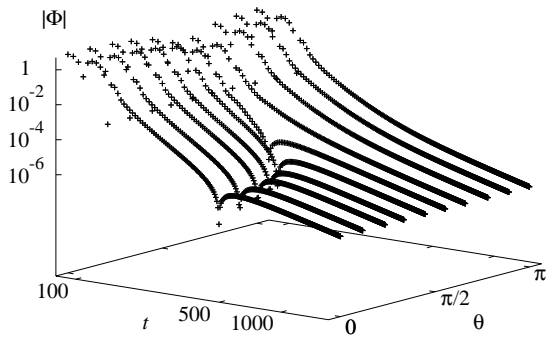}
\caption{\label{figure-2}
Plots of $ | \Phi | 
$ for $ m=0 $, $a=0.9999 $ for different angular directions $0 \leq \theta
\leq \pi/2$. The initial pulse was given by $l=2$. The time $t
$ and $ | \Phi | 
$ are shown on a logarithmic scale. For late times the numerically determined 
power-law coefficient is $\mu = 3.15$.}
\end{figure}

The main conclusion from the examples described in \cite{PaperI} can be 
stated in the following way: For a fixed value of the azimuthal separation
parameter $m$, the late time dynamics is dominated by the lowest allowed 
$l$-multipole that is compatible with the choice of $m$ ($l
$ has to satisfy the condition $l \geq m$) and with the equatorial 
symmetry properties of the initial data. Note that symmetric and antisymmetric 
modes do not get mixed during the evolution.
The fall-off behavior of the dominating multipole is given by 
$| \Phi | \propto t^{-(2l+3)}$.
This statement is illustrated in Fig.\ \ref{figure-2} by a specific example
in which the initial pulse is given by $l=2, m=0$, which is not the lowest 
axisymmetric mode that is symmetric with respect to the equatorial plane: 
During the evolution there is a transition to the $l =m=0$ mode. 
The numerically 
determined power-law coefficient governing the decay of the $l =m=0$ mode 
is $\mu = 3.15$ \cite{adPaperI}. What, in this logarithmic diagram, appears to 
be a notch for angles smaller than $\theta \approx 0.9$, in fact represents
the change of sign that the field $\Phi$ experiences for those angles during 
the transition from the initial $l=2$ distribution to the $l=0$ mode.

The example shown in Fig.\ \ref{figure-2} 
illustrates the late-time behavior for one special choice of 
parameters. However there are several questions that remain open: What happens, 
for example, if we take initial data given by an $l=4$ multipole? In analogy to
the situation depicted in Fig.\ \ref{figure-2}, we expect a transition to
the $l=0$ mode that will decay as $t^{-3}$.
More generally, is seems obvious that {\em any} axisymmetric initial pulse with 
equatorial symmetry will descend to $l=0$ and that 
the late-time power-law decay of the field $\Phi$ will be given by 
$| \Phi | \propto t^{-3}$.

Recently, these and other issues have been addressed in two different 
analytic studies of late-time phenomena occurring in the course of the 
evolution of a massless scalar field on rotating black hole backgrounds.

L.\ Barack and A.\ Ori \cite{BarackOri} have 
presented an analysis that is done
in the time domain and yields expressions for the asymptotic behavior at
future null infinity, at the black hole horizon, and at timelike infinity.
After decomposing the field $\Phi$ into spherical harmonics,
\begin{displaymath}
 \Phi = (r^{2}+a^{2})^{-1/2} \sum_{lm} 
\Psi^{lm}(t,r) Y_{l}^{m}(\theta,\phi) \; , 
\end{displaymath}
they consider an expansion of $\Psi^{lm}(t,r)$  in  inverse powers of 
advanced time (late-time expansion). 
According to their prediction, to leading order in 
$t$, the late-time decay of a mode 
specified by $l$ and $m$ at timelike infinity is given by
\begin{equation}
\label{Amospredict}
 \Phi \propto t^{-l-|m|-3-q}
 \; ,
\end{equation}
where $q=0$ if $l+m$ is even, and $q=1$ otherwise. Note that this 
prediction by Barack and Ori applies only in the case in which the 
$l=0$ mode is present in the initial data.

Independently, S.\ Hod has 
carried out an analysis that is based on a frequency
decomposition of the scalar field $\Phi$ \cite{Hod}. 
His analysis leads to the following 
conclusion for the late-time behavior at timelike infinity: 
The late-time behavior at timelike infinity is dominated by 
the lowest allowed multipole, $l=|m|$, if $l^*-|m|$ 
is even, and $l=|m|+1$
otherwise, where $l^{*
}$ denotes the initial multipole.
The late-time fall-off is given by
\begin{mathletters}
\label{Hodpredict}
\begin{eqnarray}
 \Phi & \propto & t^{-l^*-|m|-p-1} \;\;\; \mbox{if} \;\; l^* \geq |m|+2 \; ,\\
 \Phi & \propto & t^{-2 l^*-3  } \;\;\;\;\;\;\;\;\;\;\;\, 
  \mbox{if} \;\; l^* = |m|,|m|+1 \; ,
\end{eqnarray}
\end{mathletters}
where $p=0$ if $l^*-|m|$ 
is even, and $p=1$ otherwise.

For certain special choices of parameters, the expressions 
(\ref{Amospredict}) and (\ref{Hodpredict}) are in agreement.
Consider for example the case of an initial pulse given by $l=m=0$.
Barack and Ori, as well as Hod, predict a late-time fall-off given by 
$|\Phi |\propto t^{-3}$. In this case both analytic predictions also 
agree with the numerical results, shown in \cite{PaperI}.

In general, however, a comparison of 
the expressions (\ref{Amospredict}) and (\ref{Hodpredict}) is
made difficult by the fact that, as pointed out above, 
Eq.\ (\ref{Amospredict}) does not allow a general statement
about the dominating late-time behavior.
In contrast to (\ref{Amospredict}), Hod's formula (\ref{Hodpredict}) 
makes a definite statement about the dominating late time decay, depending
on the initial multipole.
If one, for example, sets $m=0$ and chooses an initial pulse with 
$l^{*}=4$, Hod's analysis implies $|\Phi |\propto t^{-5}$. According to Hod's 
analysis, the $l=0$ multipole ``remembers'' the initial configuration $l^{*}$
in the sense that its decay rate does not depend only on the choice of 
$m$ and the equatorial symmetry properties of the initial data, but also on
$l^{*}$.

\vspace{0.1cm}
The presence of two different analytic studies, one of which contains a rather 
surprising prediction, has motivated us
to carry out a numerical investigation of the late time fall-off behavior.

For the axisymmetric evolution studies presented in this paper, we use the 
numerical $(2+1)$ evolution code from \cite{PaperI} with the black hole mass 
set to unity. We made several modifications to the previously used code.
Instead of using the Boyer-Lindquist 
$\theta$ coordinate, we are using $y = - \cos \theta$, because 
this coordinate choice allows us to keep track of the behavior of the 
various angular modes more accurately.
The $y$ derivatives of $\Psi$ are approximated using fourth order finite 
differences. We have used quadruple precision floating point arithmetic for 
the numerical computations, so that we can follow the dynamics up to late 
evolution times. This issue is not of importance for low initial multipoles, but 
for $l=4$ and higher it becomes relevant because of the steep fall-off at the 
beginning of the tail regime. The accuracy 
of the code was tested using standard convergence tests, based on numerical
results obtained for short evolution times.

For $m=0$, in the $t,r^{*}, y$ coordinates, the wave equation 
for $\Psi(t,r^{*},y)$ reads 
\begin{eqnarray}
\label{2dwave}
&&\partial_{tt} \Psi - \frac{(r^2+a^2)^{2}}{\Sigma^{2}}\partial_{r^* r^*} 
           \Psi
           -\frac{\Delta (1-y^{2})}{\Sigma^{2}}\partial_{yy} \Psi \nonumber \\
&&           - \frac{2 r \Delta}{\Sigma^{2}}\partial_{r^*} \Psi
           + \frac{2 y \Delta}{\Sigma^{2}}\partial_{y} \Psi = 0 \; ,
\end{eqnarray}
where we are using the standard abbreviations $
\Delta  \equiv  r^{2
} -2Mr +a^{2}$, and
$
\Sigma^{2} \equiv  (r^{2
} + a^{2
} )^2 - a^2 \Delta (1-y^2)$ .
The initial data are given by the Gaussian-like,
initially outgoing pulse with compact support given by $50M \leq r^{*
} \leq 150M$ used in \cite{PaperI}.

The following figures represent an analysis of the data obtained from
numerical evolutions. For a range of time parameter values $t
$, the signal $\Psi$ at fixed radius is decomposed into angular 
modes. We start the evolutions with modes that are symmetric 
with respect to the equatorial plane ($l=0$ or $4$). 
Since symmetric modes do not get mixed with antisymmetric 
modes, we can discard the latter and just decompose the 
signal into even modes. Since we expect the signal at late times
to be dominated by the lowest multipoles, we only take into account the three 
lowest even modes,
\begin{eqnarray}
&&\Psi(t,r^{*}=\mbox{const.},y)= \\
&& \alpha(t) P_{0}(y) + \beta(t) P_{2}(y) + \gamma(t) P_{4}(y) 
\; , \nonumber 
\end{eqnarray}
where the $P_{0,2,4}(y)$ are the standard Legendre polynomials.

\begin{figure}
\epsfxsize=0.48\textwidth
\epsfbox{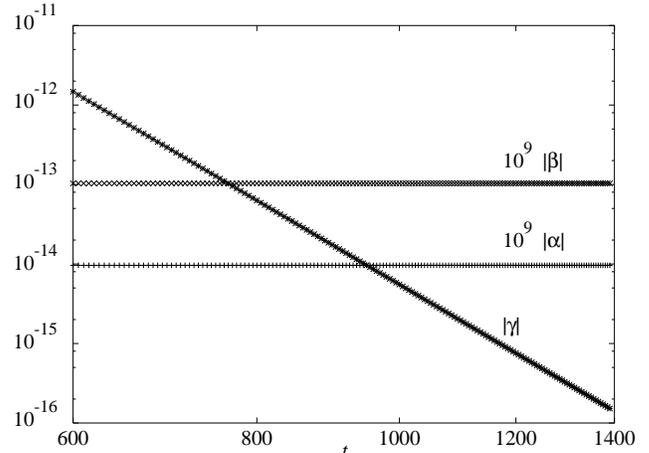}
\caption{\label{figure-3}
Log-log plots of the coefficients $\alpha$, $\beta$, 
and $\gamma$ for $ m=0 $, $a=0 $, evaluated at $r^{*}=10$.  
The initial data are given by 
an $l=4$ pulse. For late times, the decay of the $l=4$ mode 
is governed by the power-law $\gamma(t) \propto t^{-10.93}$.
}
\end{figure}

In Fig.\ \ref{figure-3} we depict the fall-off behavior of 
the absolute values of the 
expansion coefficients $|\alpha|$, $|\beta|$, and $|\gamma|$, extracted
at $r^{*}=10$,
as functions of the Boyer-Lindquist time $t$ for the test case $a=0$. 
The grid spacings are given by $dr^{*} = 0.04$ and $dy=0.01$.
The initial pulse is given by the $l=4$ multipole, and the 
computation is done with the same $(2+1)$ evolution code that is used in 
the $a \neq 0$ case. Since for $a=0$ the background
is spherically symmetric, the $l=4$ multipole has to be 
preserved during the evolution. Hence we know that all contributions from 
modes other than $l=4$ are numerical artifacts.  
A least squares fit of a function $f(t)= c \, t^{-\mu}$ 
to $\gamma(t)$ yields the power-law $\gamma(t) = -3.33 \times 10^{18}\,
t^{-10.93}$. The power-law 
coefficient $\mu$ is within less than $0.7 \%$ 
in agreement with the theoretical value of $\mu=11$.
The expansion coefficients of the numerical contamination modes $l=0,2$ 
are orders of magnitude smaller than the $l=4$ mode, hence establishing 
the accuracy of the numerical evolution method.

What can we expect for $a\neq 0$? Intuitively we might anticipate a 
transition to the $l=0$ mode that will decay as $t^{-3}$, similar to 
the situation illustrated in Fig.\ \ref{figure-2}. However we know that 
Hod's analytic prediction, which we briefly discussed above, yields 
$|\Psi| \propto t^{-5}$ for the dominating $l=0$ mode at late times, contrary 
to a naive guess. 
For grid spacings given by $dr^{*} = 0.04$ and $dy=0.01$, a fit to the 
numerically computed $\alpha(t)$ yields the power-law $\alpha(t)
\propto t^{-6.04}$ for the 
decay of the $l=0$ mode which is dominating at late times. 
This result is deeply surprising.
The transition to the $l=0$ mode is quite intuitive, but we might 
have expected a decay rate of $t^{-3}$, independently of the 
particular choice of the initial value of $l$, as long as the initial 
pulse is symmetric with respect to the equator.
The numerical result is not only 
counter-intuitive, it also seems to be 
incompatible with the analytic prediction given in 
\cite{Hod}.
This suggests that the computed $\alpha(t) \propto t^{-6.04}$ behavior might 
represent a numerical artifact rather than a physically meaningful result.
\begin{figure}
\epsfxsize=0.48\textwidth
\epsfbox{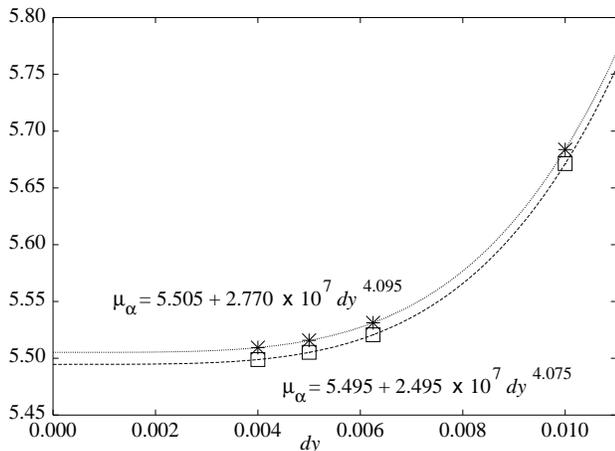}
\caption{\label{figure-4}
Plots of  $\mu_{\alpha
}$ for $ m=0 $, $a=0.9999 $, observed at $r^{*} = 10$ (boxes) and
$r^{*} = 25$ (stars). The initial pulse was given by $l=4$.
} 
\end{figure}
This suspicion is confirmed by Fig.\ \ref{figure-4}, that shows the grid size 
dependence of the numerically extracted power-law coefficient  $\mu_{\alpha}$,
that governs 
the decay of $\alpha$ at $r^{*} = 10$ (boxes) and at $r^{*} = 25$ (stars)
The results were obtained from four 
runs with $dr^{*} = 0.16,0.1,0.08,0.064 $ and 
$dy = 0.01,0.00625,0.005,0.004$, i.e.\ the discretization in the $y$ direction 
was chosen to be finer than in the cases above.
To limit the computational expenses, the run times were shorter than in the 
previously considered cases, 
and $\mu_{\alpha} $ was extracted for $600 < t < 800 $.
As illustrated in Fig.\ \ref{figure-4}, a fit of a function $\mu_{\alpha}
= c_{1} + c_{2} \, dy^{c_3}$ to the numerical data yields a value of $c_3$ 
that is close to $4$,
indicating that the 
discretization error is dominated by the finite difference
approximation of the $y$ derivatives. 
The continuum limit of $\mu_{\alpha}$, however, seems to be neither $5$ nor 
$6$ and shows a rather weak dependence on the location of the observer, 
possibly a numerical artifact. Our numerical work strongly suggests that an 
extension of the existing analytic and numerical studies is necessary.

\vspace{0.1cm}
The studies presented in this paper show that, for $m$ set to zero and 
initial data with equatorial symmetry, the late-time behavior of scalar 
test fields on Kerr backgrounds is dominated by the lowest allowed 
$l$-multipole, namely $l=0$. The numerical results imply that, in general, 
the late-time fall-off of that dominating multipole is different from that 
in the Schwarzschild case, and appears not to agree with a central result 
of Hod's recent analytic study \cite{Hod}. Compatibility of our numerical 
results with the analysis of Ori and Barack \cite{BarackOri}, requires an 
extension of their studies to initial data that do not contain the $l=0$ 
multipole. The characterization that the initial pulse is ``pure $l=4$'' is 
ineluctably related to the use of the Boyer-Lindquist $\theta$ coordinate. 
Without explicit reference to the Boyer-Lindquist coordinate
system, the numerical 
result can be stated as: There is an axisymmetric initial pulse with 
equatorial symmetry that exhibits a transition to $l=0$ and the late-time 
power-law decay of the field $\Phi$ is given by $|\Phi | \propto t^{-\mu}$, 
where the numerically determined power-law coefficient $\mu$ 
is different from the
Schwarzschild value $\mu = 3$. 
Additionally, our recent studies suggest that conclusions 
about the late-time power-law decay, made in \cite{PaperI} on the basis of
particular examples, are not valid for generic situations.

The numerical results suggest that further analytic work is needed. It will 
also be useful to perform numerical computations up to later evolution times 
to underscore the validity of the statements made in this paper. 
Future numerical work should include studies of higher, non-axisymmetric 
multipoles, and also the physically relevant case of gravitational 
perturbations.

\vspace{0.1cm}
I am grateful to Amos Ori for originally suggesting this work to me.
It is a pleasure to thank Richard H.\ Price for many encouraging 
discussions and helpful suggestions. 
I also wish to thank Pablo Laguna, Amos Ori, and Jorge Pullin for 
helpful discussions.
This work was partially supported by NSF grant PHY-9734871.
An allocation of computer time from the Center for High Performance Computing 
at the University of Utah is gratefully acknowledged. CHPC's SGI Origin 2000 
system is funded in part by the SGI Supercomputing Visualization Center 
Grant.


\begin{thebibliography}{99}
\bibitem{adPaperI} Please note that Fig.\ 6 from \cite{PaperI}
                   should exhibit the same qualitative features as 
                   Fig.\ \ref{figure-2}.
                   As a careful error 
                   analysis in 
                   connection with the work published in this paper has 
                   shown, Fig.\ 6 in \cite{PaperI}
                   does not correctly describe the transition 
                   from the $l=2 $ to the $l=0$ multipole (due to a relatively 
                   coarse grid spacing in the Boyer-Lindquist $\theta$ 
                   coordinate).
\end{thebibliography}
\end{document}